\documentclass[preprint]{revtex4}
\usepackage{graphicx}

\begin{document}
\begin{titlepage}
\title{Fragility and elastic behavior of a supercooled liquid}.\\
\author{Madhu Priya and Shankar P. Das}

\affiliation{School of Physical Sciences,
Jawaharlal Nehru University,\\
New Delhi 110067, India.}

\vspace*{1.5cm}
\begin{abstract}
A model for the supercooled liquid is considered by taking into
account its solid like properties. We focus on how the long time
dynamics is affected due to  the coupling between the slowly
decaying density fluctuations and the local displacement variables
in the frozen liquid. Results from our model agree with the recent
observation of Novikov and Sokolov [Nature (London) {\bf 431},
961 (2004)] that the fragility
index $m$ of a glass forming material is linearly related to the
corresponding ratio $K/G$ of the bulk and the shear moduli.
\end{abstract}

\vspace*{.5cm}


\maketitle
\end{titlepage}


\section{Introduction}

Understanding the basic mechanism for solidification of a
supercooled liquid into an amorphous structure constitutes an area
of much current research interest in Condensed matter physics. As
the liquid is increasingly supercooled its shear viscosity $\eta$
increases and the dynamics slows down enormously. The glass
transition temperature $T_g$ is characterized by the generic value
of $\eta \sim 10^{14}$P with the corresponding relaxation time
reaching the laboratory time scales. Experiments show that the
nature of relaxation in different supercooled liquids approaching
the glass transition is not universal. Liquids have been classified
as fragile or strong depending on their dynamical behavior in the
vicinity of $T_g$. Strong liquids show a steady increase of
viscosity with the lowering of temperature. In fragile systems the
viscosity first increases slowly in the temperature range higher
than $T_g$ and this trend is followed by a much sharper increase of
$\eta$ near $T_g$. This classification of different glass forming
materials is facilitated in terms of the so called fragility
parameter defined \cite{angell} as,
\begin{equation}
\label{kin-frag}
 m =\partial \ln \eta/\partial (T_g/T)|_{T_g}.
 \end{equation}
More recently evidence of a dynamic crossover in the value of the
viscosity induced by pressure change has also been reported
\cite{roland}. An associated characteristic of the supercooled state
is its solid like behavior \cite{dyre}. This is manifested in the
elastic response of the system to an applied stress. The elastic
behavior persists for times shorter than the structural relaxation
time. The frozen solid  with amorphous structure has transverse
sound modes in addition to the longitudinal modes which are present
in the normal liquid state. Recently Sokolov and Novikov
\cite{nature} have demonstrated that the fragility parameter $m$ of
a liquid is linearly related to the corresponding ratio $K/G$ of its
bulk and shear moduli, {\em i.e.}, $ m-17= 29(K/G -1)$. The
dependence of the fragility and the vibrational properties of the
liquid on the basic interaction potential was also tested recently
with computer simulations \cite{ngai}. This strengthens further the
possibility of understanding the vibrational and relaxation
properties of the frozen liquid from a common standpoint.

In the supercooled state  of the liquid, a tagged particle makes
rattling motion and is temporarily trapped in the cage formed by its
neighboring particles. The cage eventually breaks giving rise to
continuous particle motion. The  time for which the tagged particle
is localized grows with supercooling. The time correlation function
of the collective density fluctuations develops a plateau over
intermediate times and eventually decays to zero.  Ergodicity thus
persists in the supercooled liquid over longest time scales. This
behavior of the density correlation function is explained in the
Mode coupling theory (MCT)\cite{rmp} of supercooled liquids. The
basic theoretical scheme used in this approach involves a
formulation of the dynamics in terms of a few slow variables. In
this paper we present using the mode coupling approach, a model
which includes the solid like properties of a supercooled liquid. We
study the effects on the dynamics due to the couplings of slowly
decaying density fluctuations with the extra slow modes present in
the amorphous solid. Our results conform to the above linear
relation between $m$ and $K/G$ reported in Ref. \cite{nature}. The
paper is organized as follows. In the next section we present the
model for the dynamics using an extended set of slow modes for the
liquid state. In section III we present the results from the model
and demonstrate how it explains the observed data for simple
liquids. We end the paper with a discussion.

\section{Model Studied}
\subsection{Nonlinear dynamics of Slow modes}
The dynamics of the liquid is formulated in terms of a small set of
slow modes. The MCT is a basic step in this direction for studying
the dynamics of a strongly interacting liquid (at high density) by
taking into account the effects of coupling the slow modes in the
system. The simplest version of MCT deals with the conserved
densities of mass and momentum respectively denoted by $\rho({\bf
x},t)$ and ${\bf g} ({\bf x},t)$. We assume the system to be a
collection of $N$ classical particles each of mass $\bar{m}$.
$\vec{r}_{\alpha}(t)$ and $\vec{p_{\alpha}}(t)$ are respectively the
position and the momentum of the $\alpha$-th particle at time $t$.
For the density $\hat{\rho}$ and the momentum density
$\hat{\vec{g}}$ we use the standard prescription \cite{forster},
\noindent

\begin{eqnarray}
\label{micr} \hat{\rho} (\vec{x},t) &=& \bar{m} \sum_{\alpha=1}^{N}
\delta ( \vec{x} -\vec{r_{\alpha}}(t) )   \ , \nonumber \\
\hat{g_i} (\vec{x},t) &=&  \sum_{\alpha=1}^{N} p_{\alpha}^i (t)
\delta ( \vec{x} -\vec{r_{\alpha}}(t) )  \ .
\end{eqnarray}

\noindent For the solid like state in which the particles vibrate
about their mean positions, the above set of slow modes is further
extended to include the new collective variable \cite{MPP} ${\bf
u}({\bf x},t)$. This is defined in terms of the displacements ${\bf
u}_{\alpha}(t)$ of the $\alpha$-th particle, ($\alpha =1$ to $N$)
from their respective mean position denoted by ${\bf
R}_{\alpha}^{o}$ \cite{das-sch} such that ${\bf R}_{\alpha}(t)= {\bf
R}_{\alpha}^{o}+{\bf u}_{\alpha}(t)$. We adopt the definition,

\begin{equation}
\label{u-defn} \hat{\rho} (\vec{x},t) \hat{u_i} (\vec{x},t) =
\bar{m} \sum_{\alpha=1}^{N} u_{\alpha}^i (t) \delta ( \vec{x}
-\vec{r_{\alpha}}(t) )   \ ,
\end{equation}

\noindent The metastable positions of the atoms
 ${\bf R}_{\alpha}^{o} $ ($\alpha =1$ to $N$) in the glassy system remain
unaltered for a long time. In a crystal with long range order these
positions are independent of time. The equation of motion for ${\bf
u}$ is obtained using standard procedure\cite{ma-mazenko,das-sch} in
the form of a generalized Langevin equation,
\begin{equation}
\label{Leq-u} {{\partial u_i} \over {\partial t}}+ \frac{\bf
g}{\rho} \cdot {\bf \nabla} u_i = \frac{g_i}{\rho} -{\Gamma}_{ij}
{{\delta F} \over {\delta u_j}} +f_i
\end{equation}
\noindent where indices $i,j$ refer to the different spatial
coordinate axes. The Gaussian noise $f_i$ is related to the bare
damping matrix $\Gamma_{ij}$, through the fluctuation dissipation
relation, $<f_i({\bf x},t)f_j({\bf x'},t')> =  2k_B T {\Gamma}_{ij}
 \delta ({\bf x}-{\bf x'} ) \delta (t-t')$, where
$k_B T = {\beta}^{-1}$ is the Boltzmann factor. The average thermal

speed of a liquid particle of mass $\bar{m}$ is
$v=1/\sqrt{\beta\bar{m}}$. $F[\rho,{\bf g},{\bf u}]$ is the
effective Hamiltonian such that the probability of the equilibrium
state is given by $e^{-F}$. For the isotropic solid $F$ is obtained
in the general form \cite{bkim,das-sch},

\begin{equation}
\label{free-en} F = \int \frac{\bf dx}{2} \left [\frac{g^2}{\rho}+ A
{(\delta \rho)}^2 + 2B {\delta \rho} s_T + K s_T^2 + 2G
\tilde{s}_{ij} \tilde{s}_{ji} \right ]
\end{equation}
\noindent where ${\delta \rho}= \rho - \rho_o$ is the density
fluctuation and $\rho_o$ is the average mass density. The quantities
$A$ and $B$ in (\ref{free-en}) are related to the static structure
factor ( correlation  of  density fluctuations at equal time ) for
the amorphous solid. The symmetric strain tensor field $s_{ij}$ is
defined in terms of the gradients of the displacement field ${\bf
u}({\bf x})$, $2s_{ij} = (\nabla_i u_j + \nabla_j u_i ) - \nabla_i
u_m \nabla_j u_m$ ( such that $s_{ij}=s_{ji}$).  The trace and the
traceless parts of $s_{ij}$, respectively defined as $s_T=\sum_i
s_{ii}$ and $\tilde{s}_{ij}=s_{ij}-\delta_{ij}s_T/3$, appear in the
expression (\ref{free-en}) for free energy of the isotropic solid.

The equation of motion for $\rho$ is the continuity equation,
\begin{equation}
\label{cont-eqn} \frac{\partial\rho}{\partial{t}}  + {\bf \nabla}
\cdot {\bf g} = 0.
\end{equation}

\noindent For the momentum density ${\bf g}$ the dynamics is given
by the generalized Navier-Stokes equation,
\begin{equation}
\label{nav-stk} \frac{\partial{g_i}}{\partial{t}} + \sum_j
{\nabla}_j \sigma_{ij} = \theta_i,
\end{equation}

\noindent where $\theta_i$'s  are the gaussian noises related to the
bare or short time transport properties of the liquid \cite{boon}.
The symmetric stress tensor $\sigma_{ij}$ is obtained as a sum of
the reversible part $\sigma_{ij}^{R}$ and the irreversible
(dissipative) part $\sigma_{ij}^{D}$, such that \cite{das-sch}
\begin{equation}
\label{rev-stress} \sigma_{ij}^{R} = \frac{g_i g_j}{\rho} +
P\delta_{ij}-2Gs_{ij}+2s_{ij}[\bar{K}s_T+B{\delta \rho}]
+4Gs_{im}s_{jm}
\end{equation}
\noindent where $\bar{K}=K-2G/3$. The quantity $P$ in
(\ref{rev-stress}) is identified with pressure in conventional
hydrodynamics and is obtained as local functional of the
hydrodynamic  fluctuations,
\begin{equation}
\label{pressure} P=(A\rho_o-B){\delta\rho}+(B\rho_o-\bar{K}){s_T}+
  A\frac{\delta{\rho}^2}{2} - \bar{K}\frac{s_T^2}{2} -G {s_{lm}}{s_{ml}}.
\end{equation}
\noindent The dissipative part of the stress tensor $\sigma_{ij}^D$
is expressed in terms of the bare viscosities.
\begin{equation}
\label{dissp.}\sigma_{ij}^{D}=
-\eta_{0}[\nabla_{i}v_{j}+\nabla_{j}v_{i}-\frac{2}{3}\delta_{ij}({\bf
\nabla}.{\bf v})]-\zeta_{0}\delta_{ij}({\bf \nabla}.{\bf v})
\end{equation}
\noindent where $\eta_{0}$ and $\zeta_{0}$ are the bare shear and
bare bulk viscosities respectively and ${\bf v}\equiv {\bf g}/\rho$.

\subsection{Renormalization of transport coefficients}

The time correlation functions of the slow modes are obtained by
averaging over the noises in the nonlinear equations of motion using
standard field theoretic methods \cite{MSR}. The correlation
functions between two slow modes $\psi_\alpha$ and $\psi_\beta$ are
defined as,

\begin{equation}
G_{\alpha\beta}(12)=<\psi_{\alpha}(1)\psi_{\beta}(2)>
\end{equation}

\noindent The dispersion relations for the various hydrodynamic
modes in the system are obtained from the pole structures of the
correlation functions. The corrections to the correlation functions
due to the nonlinearities in the equations of motion for the slow
variables are expressed in terms of the self-energy $\Sigma$ defined
through the Dyson equation

\begin{equation}
\label{dyson} G^{-1}({\bf q},\omega)=[G^{0}({\bf q},\omega)]^{-1}
-\Sigma({\bf q},\omega)~.
\end{equation}

\noindent $G^{0}({\bf q},\omega)$ refers to the matrix of the
correlation functions corresponding to the  linear dynamics of the
fluctuations. Assuming the amorphous solid state to be isotropic the
correlation function can be separated in terms of a longitudinal and
a transverse part as
\begin{equation}
G_{\alpha_{i}\beta_{j}}({\bf
q},\omega)=\hat{q_{i}}\hat{q_{j}}G_{\alpha\beta}^{L}({\bf q}
,\omega)+(\delta_{ij}-\hat{q_{i}}\hat{q_{j}})G^{T}_{\alpha\beta}({\bf
q},\omega)
\end{equation}



\noindent Similarly the self-energies are split into  their
respective longitudinal and transverse parts denoted by $\Sigma_{L}$
and $\Sigma_{T}$. Using the Dyson equation (\ref{dyson})  the
renormalized viscosity of the liquid taking into account the
nonlinearities in the dynamics  is obtained in the form

\begin{equation}
\Gamma({\bf q},\omega)=\Gamma_{0}+\left[\frac{1}{2k_{B}T}\right]
\gamma_{\hat{g}\hat{g}}^{L}({\bf q},\omega)
\end{equation}

where

\begin{equation}
\Sigma_{\hat{g}\hat{g}}^{T,L}({\bf q},\omega)=-q^{2}
\gamma_{\hat{g}\hat{g}}^{T,L}({\bf q},\omega)
\end{equation}

\noindent  The functional forms of the self energies are computed
with the diagrammatic techniques of the MSR field theory \cite{DM}.
Of particular interest in the context of glassy dynamics is the
density auto correlation function $\phi(q,t)$. The Laplace transform
of $\phi(q,t)$ (which is normalized with respect to its equal time
value) is obtained as a two step continued fraction in terms of the
memory function $\Gamma(q,z)$ \cite{boon},
\begin{equation}
\label{dacf} \phi(q,z)={{1}\over{z-\Omega_o^2/[z+\Gamma(q,z)]}}~~.
\end{equation}

\noindent   $\Omega_0$ is the microscopic frequency of the liquid
state\cite{hansen}. For the case of a normal liquid with only the
standard set of conserved densities as a slow variable, the dominant
nonlinearity  in the momentum equation comes from the coupling of
the density fluctuations in the Pressure functional in
(\ref{rev-stress}). This obtains \cite{ted} the standard mode
coupling model in which memory function includes only products of
density correlation functions \cite{leth,beng}. The present
formulation involving an extended set of hydrodynamic variables
which include ${\bf u}$ ( in addition to the conserved densities)
obtains the longitudinal as well as the transverse sound modes in
the amorphous solid. Their speeds are respectively obtained as,
$c_L^2=M/\rho_o+A\rho_o-2B$ and $c_T^2=G/\rho_o$\cite{bkim}, where
$M=K+4G/3$ is the longitudinal modulus. By including the coupling of
density fluctuations $\delta \rho$ with the displacement field ${\bf
u}$ ( see eqn. (\ref{rev-stress})), a new contribution to the memory
function $\Gamma(q,t)$ is obtained. At the one loop level this is
obtained as,

\begin{equation}
\Gamma({\bf q},\omega)=\Gamma_{0}+ \Gamma^{(1)}({\bf
q},\omega)+\Gamma^{(2)}({\bf q},\omega)
\end{equation}

\noindent $\Gamma^{(1)}({\bf q},\omega)$ and $\Gamma^{(2)}({\bf
q},\omega)$ are given by the expressions

\begin{equation}
\label{gamma1}
 \Gamma^{(1)}({\bf q},\omega)=\frac{8B^{2}}{k_BT}
 \int{\frac{d^{3}{\bf p}}{(2\pi)^{3}}}\int{\frac{d\Omega}
 {2\pi}p^{2}[u^{4}G_{uu}^{L}({\bf p},\Omega)+u^{2}(1-u^{2})
 G_{uu}^{T}({\bf p},\Omega)]G_{\rho\rho}
 ({\bf q-p},\omega-\Omega)}
\end{equation}

and

\begin{equation}
\label{gamma2}
 \Gamma^{(2)}({\bf q},\omega)=\frac{A^{2}}{k_BT}
 \int{\frac{d^{3}{\bf k}}{(2\pi)^{3}}}\int{\frac{d\Omega}
 {2\pi}G_{\rho\rho}({\bf k},\omega)G_{\rho\rho}({\bf q-k},\omega-\Omega)}
\end{equation}

\noindent
The role of convective nonlinearities is
assumed to be absorbed in $\Gamma_{0}$ which is the bare part or short
 time part of the transport coefficients.
The contribution $\Gamma^{(1)}$ is obtained from the first
 diagram of fig. 1 resulting  from the coupling of the displacement field
${\bf u}$ with the density fluctuation $\delta\rho$. $\Gamma^{(2)}$
is the contribution from the second diagram in fig. 1 coming from
the coupling of density  fluctuations. In evaluating these
diagrammatic contributions we make the approximation that for time
scales ( short compared to the structural relaxations) over which
the supercooled liquid displays elastic behavior, the longitudinal
and transverse correlations of the displacement field  are frozen (
constant in time ), {\em i.e.}, $\phi_{uu}^{L,T} \sim
\delta(\omega)$. Therefore as a result of the solid like behavior
over intermediate time scales, $G_{uu}^{L}$ and $G_{uu}^{T}$ are
then obtained as

\begin{equation}
G_{uu}^{L}({\bf k},t)=\frac{k_BT}{Mk^{2}}, \ \ \ \ \
G_{uu}^{T}({\bf k},t)=\frac{k_BT}{Gk^{2}}
\end{equation}

\noindent where the $k^{-2}$ dependence of the correlation function
arises from the static structure factor for the displacement fields.
Substituting these approximate forms for the displacement
correlations and evaluating the integrals in (\ref{gamma1}) and
(\ref{gamma2}), we obtain for the density correlation function and
the memory function the simple form of coupled nonlinear integral
equations,

\begin{eqnarray}
\label{dacf1}
\phi(z) &=& {{1}\over{z-\Omega_o^2/[z+\Gamma(z)]}}~~ \\
\label{mem-func}
\Gamma(t) &=& c_{1}\phi(t)+c_{2}\phi^{2}(t)
\end{eqnarray}

\noindent where in eqns. (\ref{dacf1}) and (\ref{mem-func}) we have
dropped the wave vector dependence of $\phi$ and $\Gamma$ for
simplicity. The density correlation function is approximated in the
form $G_{\rho\rho}(q,t) = \chi_{\rho\rho}(q) \phi(t)$ with
$\chi_{\rho\rho}$ being the equal time correlation of the density
correlation function determined by the thermodynamic properties like
temperature and density. The constants $c_1$ and $c_2$ in
(\ref{mem-func}) are determined from an evaluation of the vertex
functions in  this approximation ( of wave vector independence)  as,

\begin{equation} \label{const}
c_1=\frac{8}{3} \left(
\frac{\lambda_{0}\Delta_{\sigma}}{1-\Delta_{\sigma}} \right )
[1+f(\sigma)],\ \ \ \ \ \ \ \
c_2=\frac{\lambda_o}{{(1-\Delta_\sigma)}^2},
\end{equation}

\noindent where $\Delta_\sigma = \Delta_o(1-2\sigma)/(2-2\sigma)$,
and $f(\sigma)= 2/5(1-2\sigma)$ with $\sigma$ being the Poisson's
ratio $\sigma =(3K-2G)/[2(3K+G)]$. We have used in the expressions
(\ref{const}) the definitions $\Delta_o={B^2}/(AG)$ and
$\lambda_o=(\Lambda^3/6\pi^2n_o){(v/c_L)}^2$
  in terms of parameters dependent on the thermodynamic state of the system.
The equilibrium number density of particles in the fluid is $n_o$
($\rho_o=\bar{m}n_o$). $\Lambda$ is the upper cutoff of wave vector
representing the shortest length up to which the fluctuations are
considered. In considering the mode coupling expression for the viscosity
we have ignored here the presence of very slow vacancy diffusion mode
and its coupling to density fluctuations.
Finally, it is useful to note here that we are
considering  the simple form of the model in which all processes
giving rise to ergodic behavior in the asymptotic dynamics have been
ignored.

\section{Results}
The central focus in the present analysis is the time dependence of
the density autocorrelation function $\phi(t)$. From the coupled set
of eqns. (\ref{dacf}) (inverse Laplace transformed in the time
space) and (\ref{mem-func}), we obtain a nonlinear
integro-differential equation for the dynamics of $\phi(t)$,
\begin{equation}
\label{int-dif}
\ddot{\phi}(t) + \dot{\phi}(t) + \phi(t)
+ \int_0^t ds \Gamma [\phi(t-s)]{\dot{\phi}}(s) =0
\end{equation}

\noindent
where the dots refer to derivative with respect to time.
This equation is solved numerically to obtain the behavior of the density
correlation function. It is clear that the dynamics of the density
correlation function $\phi(t)$ is driven here by the memory function
$\Gamma(t)$ which is expressed in terms of $\phi$ itself. This
constitutes  a nonlinear feed back mechanism \cite{beng,leth}
causing a dynamic transition of the liquid to a nonergodic state in
which the long time limit of the density correlation function
$\phi(t\rightarrow\infty) = f$. The quantity $f$ is termed as the
non ergodicity parameter (NEP). In the plane of $c_1$ and $c_2$ the
dynamic transition occurs along the line $c_1 = 2\sqrt{c_2} -c_2$
\cite{kim-gene}.

In fig. 2 we display the phase diagram with $c_1$ and $c_2$ showing
the ideal glass transition line. Though this dynamic transition is
finally cutoff due to ergodicity restoring mechanisms, it marks a
cross over in the dynamical behavior of supercooled liquid. The
liquid state is characterized by the density correlation function
following initial power law relaxations, followed by a final
stretched exponential decay $\phi \sim \exp[-(t/\tau)^{\beta}]$
\cite{gotze12,kim-gene}. In the glassy state the density correlation
function decays to a nonzero value equal to the non-ergodicity
parameter. In fig. 2 we have also shown the values of $c_{1}$ and
$c_{2}$ corresponding to the nonergodic states in which the dynamics
is studied in this paper (to be explained below). For presenting our
results in the following, we express time in units of
$\Gamma_o/({\rho_o}c_L^2)$ where $\Gamma_o=\zeta_o+4\eta_o/3$ is the
bare longitudinal viscosity of the liquid related to its short time
dynamics. We keep the $\lambda_o$ appearing in the expression
(\ref{const}) for the coupling constants of the memory function
fixed at a constant value (=.4) throughout the calculation. This
numerical value is reached by treating $\lambda_o$ as an adjustable
parameter here for comparison of the results of the present model
with experimental data. This essentially implies that the cutoff
$\Lambda$ of wave vector integration is being treated as an
adjustable parameter in the coarse grained model we present here.

A key quantity of interest in the present analysis is the fragility
parameter $m$ which by definition is related to the final relaxation
behavior of the supercooled liquid near $T_g$. But over this
temperature range,  the MCT approach has not been very
successfull\cite{rmp} in explaining the relaxation behavior. A
useful observation\cite{nature} in this respect is that the slope of
the viscosity vs. inverse-temperature plot in the high temperature
range can be linked with the fragility. This is justified as follows
: For very fragile systems (large $m$) the slope of the $\eta$ vs.
$T_g/T$ curve at temperatures near $T_g$ is large and hence it must
be correspondingly small at the other end of the Angell plot, {\em
i.e.}, for temperatures much higher than $T_g$. This is because the
curves for different $m$ meet at both ends on the $T_g/T$ axis. By
examining experimental data, it was pointed out \cite{nature} that
the slope of the $\eta$ vs. $T_g/T$ curve on the high temperature
side is inversely related to the fragility $m$. Since MCT is a valid
theory for the slow dynamics well above $T_g$, we have investigated
the relaxation behavior which follows from the present model in this
high temperature range. We focus on the growth of the relaxation
time $\tau$ instead of that of the viscosity $\eta$. In our model,
increasing the parameter $\Delta_o$ brings the system closer to the
ideal transition ( since it results in an increase of $\tau$ ) and
hence this parameter is treated like the inverse of temperature,
$\Delta_o \propto T_g/T$. The dependence of the relaxation time
$\tau$ on $\Delta_o$ at various (fixed) $K/G$ values is displayed in
fig. 3. For large $K/G$ the observed variation is similar to the
$\eta$ vs. $T_g/T$ curve of a fragile liquid. Although according to
MCT model the power law behavior describes the relaxation time data
in this range better, we fit $\tau$, following Ref. \cite{nature},
to an Arrhenius form with activation energy $\kappa$, {\em i.e.,}
$\tau \sim \exp(\kappa\Delta_o)$. By the property of the Angell plot
referred to above, we assume $\tilde{m} = 1/\kappa $ to be
proportional to the fragility $m$. $\Delta_o$ is proportional to
$T_g/T$ with the proportionality constant being independent of
$K/G$. Thus we assume that constants $A$ and $B$ in the expression
(\ref{free-en}) for the free energy are only functions of
temperature and the two elastic constants $K$ and $G$ are to be
treated as independent entities. Fig. 4 displays $\tilde{m}$ against
the variation of $K/G$ showing an approximately  linear behavior.
Therefore the present model agrees qualitatively with the linear
correlation predicted by Novikov and Sokolov. It should be noted
that the inverse nature of the relations between fragility and the
slope of the viscosity vs. inverse-temperature plot respectively at
the high and low temperature sides is a property of the Angell plot
itself. However the reciprocal relation  {\em i.e.} $\kappa \propto
1/m$ as used above can only be justified in a qualitative manner.
In order to demonstrate the sensitivity of the results on the values
of the parameters $\Delta_o$ and $\lambda_o$ we display
in fig 4. the $\tilde{m}$ vs $K/$G curves for
different values of $\lambda_{0}$ ranging from $\lambda_{0}=0.3$ to
$\lambda_{0}=0.8$. The nature of the curves are qualitatively similar to
 what the best fit value for $\lambda_o$ obtains.

Next we consider the dynamics on the other side of the transition
when the density correlation function freezes to a constant until
ergodicity restoring processes take over.  In the glassy state, the
NEP $f$ is estimated from the plateau which $\phi(t)$ reaches over
long time. Fig. 5 displays the variation of $f$ with the ratio
$K/G$. The nonergodic behavior and the elastic response of the
supercooled liquid as seen here are valid over initial time scales
for which the system is frozen and is solid like. However estimating
the fragility parameter $m$ directly involves computing the
temperature dependence of viscosity or relaxation time near $T_g$.
As already pointed out, the long time dynamics in the deeply
supercooled state close to $T_g$ is still beyond the scope of MCT.
Our calculation of the NEP at low temperatures here only refer to
intermediate time scales up to which the present description in
terms of the supercooled liquid is valid. Therefore in order to link
fragility $m$ with the elastic properties of the supercooled liquid
we make use of the dependence of the fragility $m$ on the NEP $f$ as
inferred from analysis of experimental data \cite{sokolov}. 
In  fig. 6 the data points and best fit curve for the experimental results
linking NEP $f$ with the fragility $m$, {\em i.e.}, $m= 167-176f$,
is shown.  The NEP values noted here are taken at the glass
transition temperature $T_g$ \cite{sokolov}. To test this empirical
relation further we compare the $\alpha_o=(1-f)/f$ vs. $m$ in the
inset of fig. 6 with another set of experimental data \cite{buchenau}
from X-ray Brillouin scattering. Similar qualitative behavior is
seen validating the link between the NEP with the fragility. The
dependence of $m$ on the ratio $K/G$ of elastic constants is then
obtained through their common dependence on the NEP $f$. The
variation of $m$ with $K/G$ is shown in fig. 7. We also display the
experimental data (from which the linear relation of Ref.
\cite{sokolov} was proposed) in the same figure, showing reasonable
agreement with the predictions of the present model. In reaching
this agreement between the theoretical model and experimental data
we adjust the temperature dependent parameter $\Delta_o$ to the
value 1.7. As argued above $\Delta_o$ is proportional to $T_g/T$ and
hence if we assume near $T_g$, $\Delta_o=c_o(T_g/T)$, then $c_o=1.7$ obtains the
best fit of the present model's predictions with results of Ref.
\cite{nature}. To demonstrate the sensitivity of the results on the
values of the adjusted parameters we  display in fig 7. ( as in fig
4 for $\tilde{m}$ ) the fragility $m$ vs $K/$G curve for different
values of $\lambda_{0}$ ranging from $\lambda_{0}=0.3$ to
$\lambda_{0}=0.8$. $\Delta_{0}$ value is kept same for all the
curves {\em i.e.} $\Delta_0=1.7$. Usually fragility index
measurements refer to the slope of the viscosity vs. inverse
temperature plot at the calorimetric glass transition temperature
$T_g$. Hence the parameter $\Delta_o$ has been kept fixed here. The
roughly linear behavior of the  $m$ vs. $K/G$ curve is retained over
the whole range of the value of the parameter $\lambda_{0}$. The
variations of $f$ and $m$ with $K/G$ as displayed in the fig. 5
and 7 respectively are obtained at this same fixed value for
$\Delta_o$. With the variation of $K/G$, the corresponding values of
$c_1$ and $c_2$ are changed as shown by filled circles in the fig.
2. With increasing value of $K/G$ we move to regions of small values
of the couplings $c_1$ and $c_2$.


Finally we make two predictions using the proposed model linking
the elastic behavior with relaxation properties in the glassy systems.
According to the present model the density correlation function  $\phi(t)$ has
a complex relaxation behavior. On the ergodic side of the transition line,
 the initial power law relaxations change over to a stretched
 exponential decay with a stretching exponent $\beta$. 
As a further illustration of the present model, we display in fig. 8 the
 variation of the stretching exponent $\beta$ with the Poisson's ratio $\sigma =(3K-2G)/[2(3K+G)]$.
For the results shown here we keep $\Delta_o$ fixed at a typical
liquid state value $\Delta_{0}=0.885$. It should be noted that following
the discussion above the $\beta$ presented here corresponds to temperatures
 much above $T_g$. On the other hand for computing the fragility $m$ vs. Poisson's
 ratio $\sigma$ in terms of the non ergodicity parameter $f$, as described
 above in Fig. 5-7, we work on the nonergodic side of the transition line
, {\em i.e.} in the
glassy phase where the freezing of the density correlation function
is  clearly visible ($\Delta_o=1.7$). We do not have in the present
mode coupling model a way of computing the $\beta$ in this low temperature regime
which will require taking into account the role of ergodicity restoring
mechanisms in the long time dynamics. Hence the predicted behavior of the
stretching exponent should be tested at higher temperatures away from
$T_g$.

Another suitable quantity is the power law exponent of relaxation near
$T_g$. On the nonergodic side of transition line,
 the system is dynamically frozen 
initially until the ergodicity restoring processes take over.
 For initial time scales ( beyond the microscopic times)
the density correlation function develops a
plateau $f$ with a power law form $\phi(t)=f+Ct^{-a}$ decay. From the
present model we are able to predict the dependence of the power law
 exponent $a$ on the $K/G$ ratio. Fig. 9
displays the dependence of the exponent $a$ on $K/G$. The inset
shows the nature of the power law behavior at a particular value of
$K/G=2.6$. The curve is obtained at $\Delta_{0}=1.7$.
 According to the predictions of the present model the
power law exponent grows sharply for less fragile systems indicating a
sharper freezing of the system over intermediate time scales for strong
liquids.

\section{Discussion}

The present model for the amorphous system is constructed in terms
of an extended set of slow modes. The approximations which go into
the present formulation of the dynamics for the solid like state are
 well known and standard. In order to define the slow variable
which accounts for the solid like nature of the amorphous state, we
require reference to a rigid lattice. On the other hand the
ergodicity restoring process in the system invalidates the existence
of any such rigid structure over longest time scales. Therefore the
elastic description of the liquid is only valid up to the time scale
of structural relaxation. A self-consistent treatment with dynamic
connections between the elastic and viscous behavior of the system
will possibly provide a way to make a unified description including
both of the above two aspects of the glassy state. Furthermore, the
atoms in the solid vibrate around the random lattice sites over
intermediate time scales. Generally strong glassy systems have a
higher network forming tendency than the highly fragile systems
\cite{physicaB}. In the latter structural degradation occurs more
easily. Thus for systems with very high $m$ values the underlying
assumptions of the present model hold for shorter time scales.

The present model which considers the relation of the fragility of
the liquid to its elastic properties brings out some basic characteristics
of the dynamics. In fig. 7 we se that the increase in the fragility $m$
corresponds to higher values of the ratio $K/G$  while from fig. 2
it follows that coupling $c_1$ and $c_2$ between the slow modes is
weaker for larger values of this ratio. Therefore the trend of the
experimental data indicates that in a more fragile system there is
a weaker coupling of the hydrodynamic modes. This is particularly
relevant since the coupling between the slow modes constitutes
( through the formulation of the mode coupling theories )
the basic mechanism for slow dynamics in supercooled
liquids.  Also, since it follows from the present theory that
larger values of $K/G$  give rise to lower values
of the nonergodicity parameter, the strength of the freezing or jamming
of the system is weaker in more fragile systems.


It is important to note that the present model is formulated
for the case of simple liquids and accordingly it is expected to
apply to only specific types of systems.  The relation between $m$
and $K/G$ stated by Sokolov and Novikov is based on the data
analysis of non metallic and non -polymeric systems\cite{prb}.
Efforts to apply this scaling to a wider group including metallic
glasses or multi component systems fail \cite{other-nature} for
obvious reasons. Polymeric materials also clearly deviate from this
linear relation.  Monomers follow similar correlations, but increase
in the chain length in some polymers leads to the deviations
\cite{ding}.


The model similar to the one used here having the memory kernel with
both linear and quadratic terms has been initially proposed in Ref.
\cite{gotze12}. However the present form of the $\Gamma(t)$ ( having
direct relation to the elastic constants) is only justified by
including the displacement variable for the amorphous solid
\cite{bkim} in the theoretical description. In this regard it is
useful to note that the past literature on MCT can be broadly
divided into two different groups. In one approach the dynamics of a
supercooled liquid is studied starting from the basic equations
which apply to the liquid state or crystalline state of matter. This
is the scheme we have followed in our analysis here. The set of
nonlinear equations we have considered for constructing the mode
coupling equations in the present context are a) the conservation
laws for the mass and momentum and b) the dynamical equation for the
extra slow modes for the solid like state. The resulting model
linking the non ergodic behavior to the elastic properties then
follows through a careful consideration of the nonlinearities in the
dynamics. The present model thus extends the standard MCT in a
physically relevant manner to include the solid like nature of the
supercooled state. On the other hand the MCT literature also have
schematic models in which ad-hoc forms of memory functions are
simply assumed in order to consider different types of possible
relaxation scenarios which follow from the nonlinear feed back
mechanism. It is only in this respect the present model is
technically similar to the schematic $\phi_{12}$ MCT studied by
G\H{o}tze. Finally, we have ignored here the wave vector dependence
of the model to focus on the basic feed back mechanism. The
parameter $\Delta_o$ has been treated as a temperature dependent
 parameter ( $\sim T_g/T$ ) in the model.
The temperature and density dependence of the results should follow
in a natural way when such extensions of the model are considered.
This calculation will require implementing the proper structure
factor of the liquid in the formulation and will be considered
elsewhere.

\section{Acknowledgement}
MP and SPD acknowledges CSIR, India for financial support.

\newpage
\begin{figure}
\includegraphics*[width=8cm]{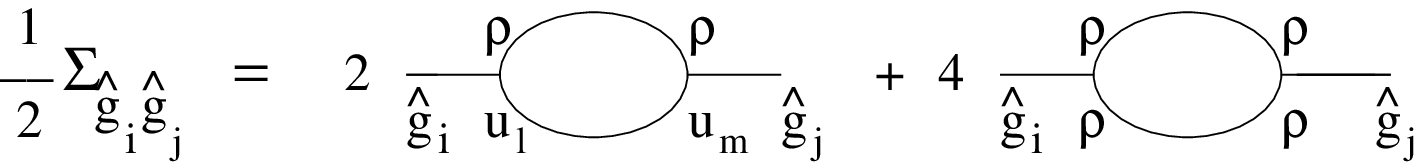}
\caption{ Relevent one loop diagrams which contribute
to the effective viscosity as non-linear corrections.}
\end{figure}

\begin{figure}
\includegraphics*[width=8cm]{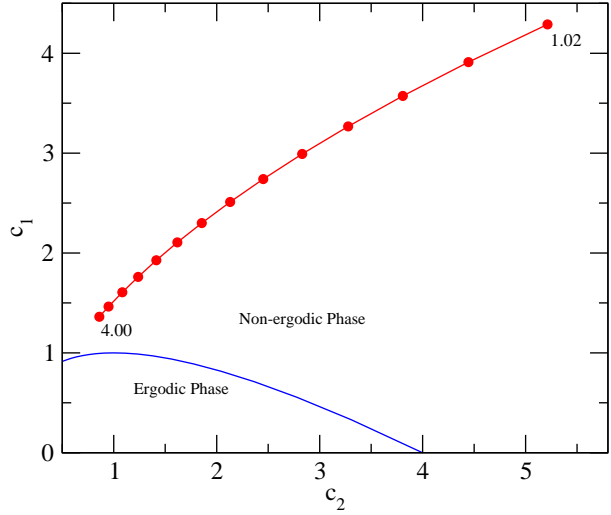}
\caption{ $c_{1}$ vs. $c_{2}$ and the solid line
displays the dynamic transition line. The circles (joined with solid
line to show continuity) shown in the nonergodic phase correspond to
the states for which the results of fig. 5
and 7 are obtained ($\Delta_{0}=1.7$).
 These circles correspond to different values of
the ratio $K/G$ of bulk and shear moduli with the adjacent number to 
a circle indicating the corresponding value of K/G.}
\end{figure}

\begin{figure}
\includegraphics*[width=8cm]{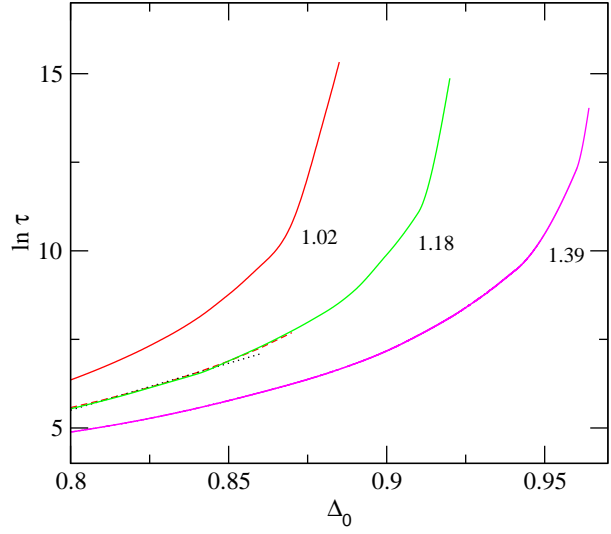}
\caption{ Log of relaxation time $\tau$ in units of
$\Gamma_0/(\rho_o c_L^2)$, vs $\Delta_{0}$ (see text). The constant
 $K/G$ value  corresponding to each curve is also displayed
 adjacent to it in the figure.
The dashed and dotted lines are, respectively, the power law and exponential
 fits to the corresponding curve.}
\end{figure}

\begin{figure}
\includegraphics*[width=8cm]{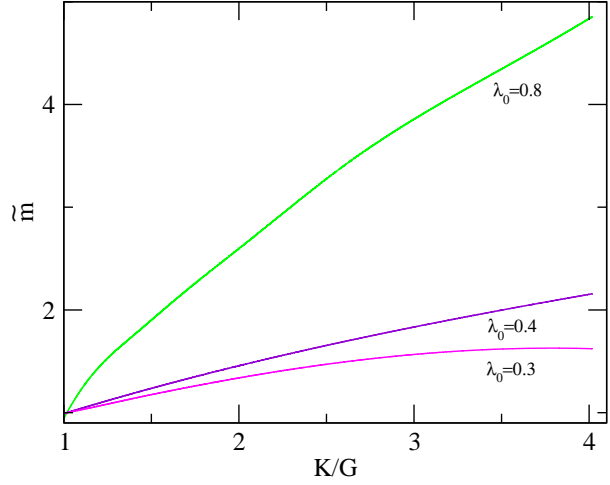}
\caption { The inverse slope $\tilde{m}$ (see text) vs
ratio $K/G$ of bulk and shear moduli at different $\lambda_{0}$ values,
displayed in the figure.}
\end{figure}

\begin{figure}
\includegraphics*[width=8cm]{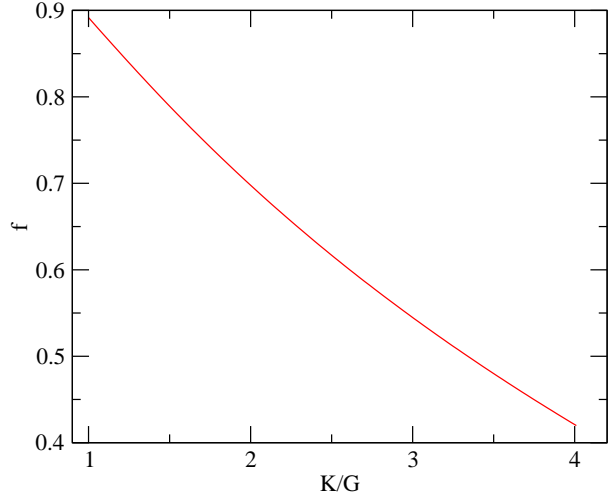}
\caption{ Non ergodicity parameter {\it f} vs the
ratio $K/G$ of bulk and shear moduli  from the present model ($\Delta_{0}=1.7$).}
\end{figure}

\begin{figure}
\includegraphics*[width=8cm]{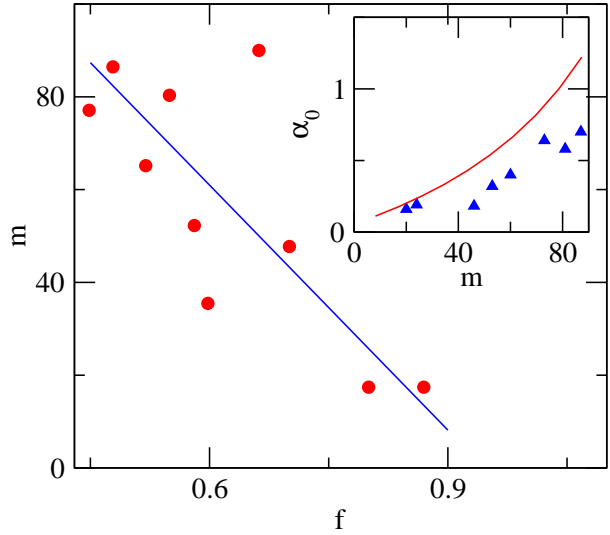}
\caption{ Experimental data (points) and best fit
(line) to the experimental data  Ref. \cite{sokolov} linking
fragility index {\it m} with $f$. The inset shows $\alpha_o=(1-f)/f$
vs fragility index $m$( solid line) (using the experimental best fit), X-ray
Brillouin scattering data ( solid triangles) \cite{buchenau}.}
\end{figure}

\begin{figure}
\includegraphics*[width=8cm]{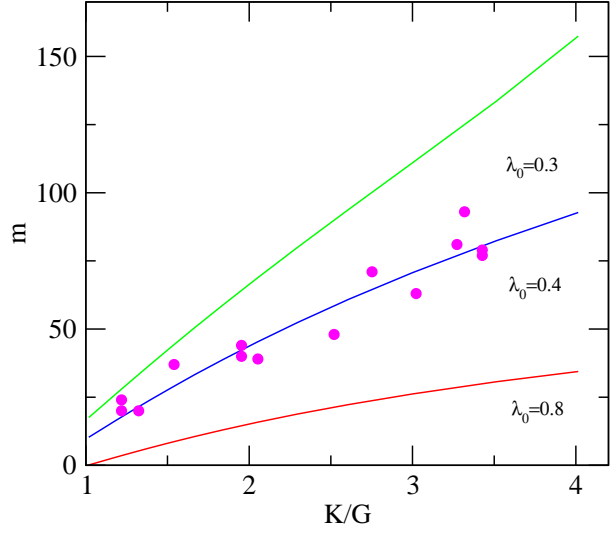}
\caption{ Fragility index $m$ vs ratio $K/G$ of bulk
and shear moduli from the present model at different $\lambda_{0}$
values (solid lines) ($\Delta_{0}=1.7$), experimental
data of Ref. \cite{nature}(points).}
\end{figure}

\begin{figure}
\includegraphics*[width=8cm]{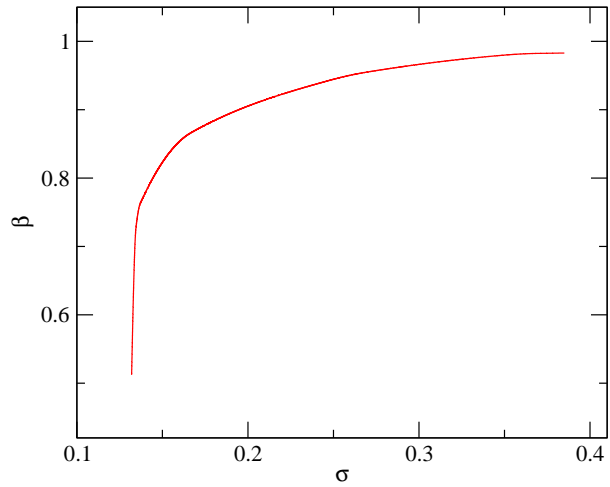}
\caption{ Stretched exponential exponent $
\beta$ vs Poisson's ratio $\sigma$ ($\Delta_{0}=0.885$).}
\end{figure}

\begin{figure}
\includegraphics*[width=8cm]{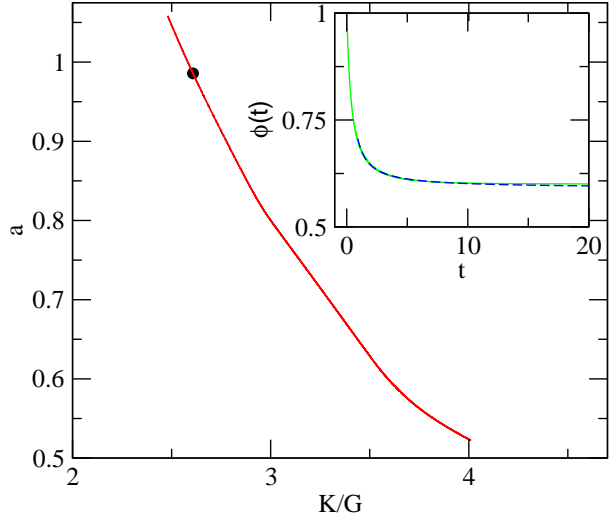}
\caption{ Power law exponent {\it a} vs the ratio
$K/G$ of bulk and shear moduli ($\Delta_{0}=1.7$).
 Inset shows normalized density
correlation function $\phi(t)$ vs {\it t} in units of
${\Gamma_{0}/(\rho_o c_L^2})$(see text). The dashed curve is the
power law fit with the exponent value shown as a dot on the main
figure.}
\end{figure}


\begin{thebibliography}{0}



\bibitem{angell}
R. B\"ohmer, K. L. Ngai, C. A. Angell, and D. J. Plazek,
 J. Chem. Phys. {\bf 99}, 4201 (1993).

\bibitem{roland}R. Casalini, and C. M. Roland, Phys. Rev. Lett. {\bf 92},
 245702 (2004).

\bibitem{dyre}J. C. Dyre, Rev. Mod. Phys. {\bf 78}, 953 (2006).

\bibitem{nature}V. N. Novikov, and A. P. Sokolov, Nature {\bf 431}, 961 (2004).

\bibitem{ngai}
P. Bordat, F. Affouard, M. Descamps, and
 K. L. Ngai, Phys. Rev. Lett. {\bf 93}, 105502 (2004).

\bibitem{rmp}S. P. Das, Rev. Mod. Phys. {\bf 76}, 785 (2004).

\bibitem{forster}D. Forster, {\it Hydrodynamic Fluctuations, Broken Symmetry,
and Correlation Functions} (Benjamin, Reading, Mass., 1975).

\bibitem{MPP}P. C. Martin, O. Parodi, and P. S. Pershan, Phys. Rev. A
{\bf 6}, 2401 (1972); P. D. Fleming, III, and C. Cohen,
 Phys. Rev. B {\bf 13}, 500 (1976).


\bibitem{das-sch}S. P. Das, and R. Schilling, Phys. Rev. E
 {\bf 50}, 1265 (1994).

\bibitem{ma-mazenko}S.K. Ma, and G. F. Mazenko,
Phys. Rev. B {\bf 11}, 4077 (1975).


\bibitem{bkim}B. Kim, Phys. Rev. A {\bf 46}, 1992 (1992).

\bibitem{boon}J. P. Boon and S. Yip, {\it Molecular Hydrodynamics}
(Dover, New York, 1991).

\bibitem{MSR}P. C. Martin, E. D. Siggia, and H. A. Rose,
 Phys. Rev. A {\bf 8}, 423 (1973).

\bibitem{DM}S. P. Das, and G. F. Mazenko, Phys. Rev. A {\bf 34}, 2265 (1986)
\bibitem{hansen}J. P. Hansen, and J. R. McDonald,
{\it Theory of Simple Liquids}, 3rd ed. (Academic, London, 2005).

\bibitem{ted}T. R. Kirkpatrick, and J. C. Nieuwoudt,
Phys. Rev. A {\bf 33}, 2651 (1986); S. P. Das, Phys. Rev. A {\bf
42}, 6116 (1990).

\bibitem{leth}E. Leutheusser, Phys. Rev. A {\bf 29}, 2765 (1984).

\bibitem{beng}U. Bengtzelius, W. G\"otze, and A. Sj\"olander,
 J. Phys. C {\bf 17}, 5915 (1984).

\bibitem{kim-gene}B. Kim, and G. F. Mazenko,
Phys. Rev. A {\bf 45}, 2393 (1992).

\bibitem{gotze12}W. G\"otze, and L. Sj\"ogren, Rep. Prog. Phys.
 {\bf 55}, 241 (1992).

\bibitem{sokolov}V. N. Novikov, Y. Ding, and A. P. Sokolov,
Phys. Rev. E {\bf 71}, 061501 (2005).

\bibitem{buchenau}U. Buchenau, and
 A. Wischnewski, Phys. Rev. B {\bf 70}, 092201 (2004) and the references cited therein.

\bibitem{physicaB}L. A. Bove, C. Petrillo, A. Fontana, A. Ivanov, C. Dreyfus, and A. P. Sokolov, Physica B {\bf 385-386}, 16 (2006).

\bibitem{prb}V. N. Novikov, and A. P. Sokolov, Phys. Rev. B {\bf 74}, 064203 (2006).

\bibitem{other-nature}S. N. Yannopoulos, and G. P. Johari, Nature {\bf 442}, E7-E8 (2006).

\bibitem{ding}A. P. Sokolov, V. N. Novikov, and Y. Ding, J. Phys.: Condens. Matter {\bf 19}, 205116 (2007).



\end{thebibliography}
\end{document}